\documentclass[twocolumn]{revtex4}
\usepackage{amssymb}
\usepackage{amsmath}

\begin{document}

\title{Uninformed Hawking Radiation}
\author{I. Sakalli$^{\ast}$}
\author{A. Ovgun$^{\ddag}$}
\affiliation{Department of Physics, Eastern Mediterranean University,}
\affiliation{G. Magusa, North Cyprus, Mersin-10, Turkey}
\affiliation{$^{\ast}$izzet.sakalli@emu.edu.tr}
\affiliation{$^{\dagger}$ali.ovgun@emu.edu.tr}

\begin{abstract}
We show in detail that the Parikh-Wilczek tunneling method (PWTM), which was
designed for resolving the information loss problem in Hawking radiation
(HR)\ fails whenever the radiation occurs from an isothermal process. The
PWTM aims to produce a non-thermal HR which adumbrates the resolution of \
the problem of unitarity in quantum mechanics (QM), and consequently the
entropy (or information) conservation problem. The effectiveness of the
method has been satisfactorily tested on numerous black holes (BHs).
However, it has been shown that the isothermal HR, which results from the
emission of the uncharged particles of the linear dilaton BH (LDBH)
described in the Einstein-Maxwell-Dilaton (EMD) theory, the PWTM has
vulnerability in having non-thermal radiation. In particular, \ we consider
Painlev\'{e}-Gullstrand coordinates (PGCs) and isotropic coordinates (ICs)
in order to prove the aformentioned failure in the PWTM. While carrying out
calculations in the ICs, we also highlight the effect of the refractive
index on the null geodesics.

\end{abstract}

\maketitle

\section{Introduction}

As is well-known, Hawking \cite{Hawking} theoretically proved that BHs could
emit radiation (often called HR), which implies that a BH would eventually
evaporate away, leaving nothing over time. This connotes a problem for QM,
which states that nothing, including information, can ever be lost. If a BH
stores whole information in its singularity forever, there would be a
fundamental flaw with QM. This phenomenon is called the information loss
paradox (a reader may refer to \cite{ILP} for the topical review). Among the
many attempts at a resolution of this problem, the most promising one came
at the turn of this century, belongs to Parikh and Wilczek (PW) \cite{PW}.
The theorem states that when a virtual pair is created just inside the BH horizon, the positive 
energy particle (real particle) can tunnel out the BH horizon by a process similar to the QM tunneling, 
whereas the negative energy particle (antiparticle) continues to stay in the BH. Conversely, as one would expect from particle-antiparticle symmetry, if a virtual pair is created just inside the horizon, the antiparticle can tunnel inward, while the real particle will eventually escape to spatial infinity.

In the PWTM, the conservation of energy is enforced. Therefore, the mass of
the BH must continuously decrease while it radiates. Besides this, the
information-carrying particle is modelled as a thin spherical shell with
energy $\omega $. Those shells could tunnel through the potential barrier,
following the principles of QM. In short, the whole tunneling process is
considered semiclassically, and the transmission coefficient is determined
by the classical action of the particle with the aid of the
Wentzel-Kramers-Brillouin (WKB) method \cite{WKB}. As a result, the obtained
spectrum is not precisely thermal, and this also leads to the unitarity of
the underlying quantum theory and the conservation of information \cite%
{Parikh}. On the other hand, so far, the solution of the PWTM to the problem
of the information paradox has not convinced everyone, and hence it has also
remained debatable (one can see the extensive review on the PWTM analysis 
\cite{Vanzo} and references therein). Furthermore, the PWTM has also
extended to the HR analysis of the non-asymptotically flat (NAF) BHs (see
for instance \cite{Park,QQJ,Sakalli1}).

The PWTM through the quantum horizon of a LDBH geometry, which is the solution to the
EMD theory \cite{Clement,Fabris,Marques}, and its extended theories \cite{MSH}, was studied in \cite{Sakalli1,Sakalli11,Sakalli2}. This BH is a NAF, four dimensional, spherically symmetric and static
dilatonic spacetime. It was shown by \cite{Sakalli11,Sakalli2} that in the proposed PW setup, there
is no correlation between different subsequently emitted particles, which
reflects the fact that information does not come out continuously during the
evaporation process. Then some possible scenarios to conserve the
information were given in \cite{Sakalli11}. To this end, the back
reaction effects were taken into account. However, we believe that, in
those studies \cite{Sakalli11,Sakalli2}, the main point that the PWTM does not yield non-thermal
radiation for a BH\ evaporating isothermally, has not been stressed enough.
Therefore, the fundamental motivation of the present study is to highlight
that the PWTM can not be the general procedure for having non-thermal HR.

In this paper, in addition to the PGCs, we also employ the PWTM within the
ICs that has not been studied before for the LDBHs. In particular, in the IC
system we represent in detail how the Hawking temperature can be precisely
obtained within the framework of the PWTM, and how the PWTM is ineffective
in achieving non-thermal radiation.

\section{Pure thermal radiation of the LDBH}

The action of the EMD in 3+1 dimensions ($4D$)\ is given by

\begin{equation}
S=\frac{1}{16\pi }\int d^{4}x\sqrt{-g}\left( R-2\partial _{\mu }\phi
\partial ^{\mu }\phi -e^{-2\beta \phi }F^{2}\right)  \label{1n}
\end{equation}

where $\phi $ is the dilatonic field with a coupling constant $\beta $ and $%
F^{2}=F_{\mu \nu }F^{_{\mu \nu }}$ in which $F_{\mu \nu }$ is the
electromagnetic field or the $U(1)$ gauge field. Static, spherically
symmetric NAF\ solutions in $4D$ were obtained in \cite{ClemLey}. Among them
the LDBH \cite{Clement}, which corresponds to the case of $\beta =1$ is
given by

\begin{equation}
ds^{2}=-fdt^{2}+\frac{1}{f}dr^{2}+R^{2}(d\theta ^{2}+\sin ^{2}\theta
d\varphi ^{2}),  \label{2n}
\end{equation}

where the metric functions and the fields are given by

\begin{equation}
\text{ }f=r_{0}^{-1}(r-b),\text{ \ \ \ \ \ }R^{2}=rr_{0},  \label{3n}
\end{equation}

\begin{equation}
e^{2\phi }=\frac{r}{r_{0}},\text{ \ \ \ \ \ }F_{rt}=\frac{Q}{r_{0}^{2}},
\label{4n}
\end{equation}

in which $b$ represents the event horizon $r_{h},$ which is also related to
the mass. In general, mass of a NAF BH is computed via the Brown-York
quasilocal mass definition \cite{BrownYork}. Thus, one can compute the
quasilocal mass of the LDBH as%
\begin{equation}
M=\frac{b}{4}.  \label{5n}
\end{equation}

Furthermore, the another parameter $r_{0}$ is related with the charge $Q$ of
the LDBH\ through

\begin{equation}
r_{0}=\sqrt{2}Q.  \label{6n}
\end{equation}

It is worth noting that both $b$ and $r_{0}$ parameters have the same
dimension in the geometrized unit system \cite{Wald} since the mass and the
charge are represented by the $[L]$ geometrical dimension .The conventional
definition of the Hawking temperature $T_{H}$ \cite{Wald} is formulated in
terms of the surface gravity $\kappa $ as $T_{H}=\frac{\kappa }{2\pi }$. For
the metric (2), $T_{H}$ becomes

\begin{equation}
T_{H}=\frac{\kappa }{2\pi }=\left. \frac{\partial _{r}f}{4\pi }\right\vert
_{r=r_{h}},  \label{7n}
\end{equation}

which yields

\begin{equation}
T_{H}=\frac{1}{4\pi r_{0}}.  \label{8n}
\end{equation}

It is clear that the obtained temperature is independent of mass, and
consequently it is constant. Therefore, $\Delta T_{H}=0,$ which means that
the radiation is an isothermal process. Thus, HR of the LDBH is such a
special radiation that the energy transfer out of it happens at a particular
slow rate so that thermal equilibrium is always satisfied. Furthermore, 
the extreme LDBH ($M$=$b$=0) is still a BH and possesses a clashed \textit{singularity 
- pointlike horizon} structure. Its singularity is null, and the delivery time for
 a emitted signal from the horizon to an external observer is infinite \cite{Fabris,Marques}. 
This extreme BH can be used to describe the LDBH remnant \cite{Sakalli11}. Using the 
massless Klein-Gordon equation, it is proven that such a remnant cannot radiate, as expected, 
and its Hawking temperature is zero. By taking the tunneling formalism with subsequent emissions and quantum gravity 
corrected entropy into consideration, Sakalli et al. \cite{Sakalli11} also showed that the entropy of the extreme LDBH can be derived.

In order to employ the PWTM and investigate the Hawking temperature of the
LDBH, one should choose a suitable coordinate system which is not singular
at the event horizon. Along the line of PW \cite{PW}, we firstly consider
the PGCs \cite{Painleve,Gullstrand} by applying the following coordinate
transformation 
\begin{equation}
dT=dt+\frac{\sqrt{1-f}}{f}dr,  \label{9n}
\end{equation}%
where the coordinate $T$ denotes the time in the PGCs, which measures the
proper time. Thus the line-element (1) transforms into 
\begin{equation}
ds^{2}=-fdT^{2}+2\sqrt{1-f}dTdr+dr^{2}+R^{2}(d\theta ^{2}+\sin ^{2}\theta
d\varphi ^{2}).  \label{10}
\end{equation}

At $r=r_{h}$ (i.e., $f=0$), the metric coefficients are all regular, and
indeed the coordinates are all well-behaved there. Since we think the
particle as an spherical shell, during the tunneling process, the particle
does not have motion in ($\theta ,\varphi $)-directions. Thus, the radial
null geodesics can be obtained as 
\begin{equation}
\dot{r}=\frac{dr}{dT}=\pm 1-\sqrt{1-r_{0}^{-1}(r-4M)},  \label{11n}
\end{equation}%
by which the upper (lower) sign corresponds to the outgoing (ingoing)
geodesics. In \cite{Hartle}, it was shown that the ratio of emission and
absorption probabilities for energy $E$ is

\begin{equation}
\frac{P_{emission}}{P_{absorbtion}}=e^{-\frac{E}{T_{H}}}.  \label{12n}
\end{equation}

In the WKB approximation \cite{WKB}, these probabilities are related to the
outgoing/ingoing imaginary part of the particle's action ($ImS_{out}/ ImS_{in}$) as follows

\begin{equation}
P_{emission}=e^{-2ImS_{out}}, \ \ \ \ \ \ \ %
P_{absorbtion}=e^{-2ImS_{in}}.  \label{13n}
\end{equation}

Since the tunnelling ratio is expressed as

\begin{equation}
\Gamma =\frac{P_{emission}}{P_{absorbtion}}=e^{-\frac{E}{T_{H}}}=e^{-2ImS},  \label{14n}
\end{equation}

where $ImS$ denotes the net imaginary part of particle's action \cite%
{Wang}. Thus, we have

\begin{equation}
ImS=ImS_{out}-ImS_{in}.  \label{15n}
\end{equation}

Meanwhile, the imaginary part of the action for the ingoing particle is
given by 
\begin{equation}
ImS_{in}=Im\int_{r_{in}}^{r_{out}}p_{r}dr
=Im\int_{r_{in}}^{r_{out}}\int_{0}^{p_{r}}dp_{r}^{\prime }dr,  \label{16n}
\end{equation}%
where $p_{r}$ denotes the canonical momentum along $r$-direction \cite{PW}. $%
r_{in}$ and $r_{out}$ represent the radial distance of the event horizon
before and after the HR, respectively. Since the BH shrinks in the process
of the HR, $r_{in}>r_{out}.$ According to the PWTM, we should fix the total
mass of the system ($M$) and allow the BH to fluctuate. Also, we consider
the chargeless particle as a thin spherical shell of energy $\omega $. After
taking into account the self-gravitational effect, mass of the BH$\ $%
decreases as $M\rightarrow $ $M-\omega $. Furthermore, Hamilton's equation $%
\dot{r}=\frac{dH}{dp_{r}}$ can be used to transform variables from momentum
to energy. Thus Eq. (16) becomes 
\begin{eqnarray}
ImS_{in}=Im\int_{r_{in}}^{r_{out}}\int_{M}^{M-\omega }\frac{dr%
}{\dot{r}}dH.  \label{17n}
\end{eqnarray}
Then, we can switch integration variables from $H$ to the particle's energy $%
\omega $. Letting $H=M-\omega ^{\prime }$, we consequently get $dH=-d\omega
^{\prime }$. So, we have 
\begin{eqnarray}
ImS_{in}=Im\int_{r_{in}}^{r_{out}}\int_{0}^{\omega }\frac{dr%
}{\dot{r}}(-d\omega ^{\prime }), 
\nonumber \\
\ =Im\int_{0}^{\omega }\int_{r_{in}}^{r_{out}}\frac{dr}{1+\sqrt{%
1-r_{0}^{-1}\left[ r-4(M-\omega ^{\prime })\right] }}\left( d\omega ^{\prime
}\right), 
\nonumber \\
\ =Im\int_{0}^{\omega }\int_{r_{in}}^{r_{out}}\frac{\Psi _{in}}{%
r-4(M-\omega ^{\prime })}dr\left( d\omega ^{\prime }\right),  
\label{18n}
\end{eqnarray}

\bigskip where

\begin{equation}
\Psi _{in}=r_{0}-\sqrt{r_{0}\left[ r_{0}-r+4(M-\omega ^{\prime })\right] }.
\label{19n}
\end{equation}

From Eq. (18), one can see that there is a contour integral in the
complexified $r$-plane picks up a residue at $r=4(M-\omega ^{\prime })$.
After deforming the contour around the pole (pushing the pole into the upper
half complex $r$-plane), we get a prefactor of $-i\pi $. For the detailed
description of residue calculus, one may refer to \cite{Mathews}. Evaluating
the integral, we obtain 
\begin{equation}
ImS_{in}=0.  \label{20n}
\end{equation}

If we repeat the same procedure for the imaginary part of the action for the
outgoing particle, we have

\begin{equation}
ImS_{out}=-Im\int_{0}^{\omega }\int_{r_{in}}^{r_{out}}\frac{%
\Psi _{out}}{r-4(M-\omega ^{\prime })}drd\omega ^{\prime },  \label{21n}
\end{equation}

in which

\begin{equation}
\Psi _{out}=r_{0}+\sqrt{r_{0}\left[ r_{0}-r+4(M-\omega ^{\prime })\right] },
\label{22n}
\end{equation}

$r$-integral seen in Eq. (21), has also a single pole at $r=4(M-\omega
^{\prime })$. Therefore, one can get

\begin{equation}
ImS_{out}=ImS=2\pi \omega r_{0}.  \label{23n}
\end{equation}

The tunneling rate (14) for a particle outwards through the horizon thus
turns out to be

\begin{equation}
\Gamma =\exp (-4\pi \omega r_{0}).  \label{24n}
\end{equation}

\bigskip So the obtained temperature

\begin{equation}
T=\frac{1}{4\pi r_{0}},  \label{25n}
\end{equation}

is nothing but the standard Hawking temperature given in Eq. (8). However
there is an intriguing issue in this result: Although the energy
conservation is enforced, the spectrum of the radiation is still precisely
thermal. According to our knowledge, this (isothermal HR) is a unique case
for the PWTM that it could not modify the pure thermal character of the HR.

Now, we want to verify our result in another regular coordinate system. For
this purpose, we consider the LDBH\ within the IC system. The ICs have
several interesting features similar to the PGCs: The time direction is a Killing vector and 
Landau's condition of the coordinate clock
synchronization \cite{Landau} is automatically satisfied. The LDBH spacetime
in the ICs has been recently studied by Sakalli and Mirekhtiary \cite{SM}.
By following the associated transformation given in that reference

\begin{equation}
r=\frac{1}{4\rho }\left( \rho +b\right) ^{2}.  \label{26n}
\end{equation}

we can express the LDBH metric in the ICs as

\begin{equation}
ds^{2}=-Fdt^{2}+G[d\rho ^{2}+\rho ^{2}(d\theta ^{2}+\sin ^{2}\theta d\varphi
^{2})],  \label{27n}
\end{equation}

with

\begin{equation}
F=\frac{1}{4\rho r_{0}}\left( \rho -b\right) ^{2},\text{ \ \ \ \ \ }G=\frac{%
r_{0}}{4\rho ^{3}}\left( \rho +b\right) ^{2}.  \label{28n}
\end{equation}

Meanwhile, the event horizon in the IC is located at $\rho _{h}=b$. From
metric (27), one can obtain the radial null geodesics as

\begin{align}
\dot{\rho}& =\frac{d\rho }{dt}=\pm \sqrt{\frac{F}{G}}=\pm \frac{1}{n}, 
\notag \\
& =\pm \frac{\rho (\rho -b)}{r_{0}(\rho +b)},  \label{29nn}
\end{align}

where $n$ is the refractive index of the medium of the LDBH geometry \cite%
{SM} and it is deterministic parameter on the imaginary part of the action
for an outgoing (tunneling) particle:

\begin{eqnarray}
\textit{Im}S_{out}=\textit{Im}\int_{\rho _{in}}^{\rho _{out}}\int_{0}^{\omega
}nd\rho (-d\omega ^{\prime }),  
\nonumber \\
\ =-r_{0}\textit{Im}\int_{0}^{\omega }\int_{z_{in}}^{z_{out}}\frac{\left[ \rho
+4(M-\omega ^{\prime })\right] }{\left[ (\rho -4(M-\omega ^{\prime })\right] 
}\frac{d\rho }{\rho }(d\omega ^{\prime }).  
\label{30n}
\end{eqnarray}

The $\rho $-integral has a pole at $4(M-\omega ^{\prime })$. However, one
must be cautious about a subtle point, which was pointed out in \cite%
{SM,Akhmedov,Chatterjee} that when one deforms the contour the integral
around the pole, the semicircular contour in Eq. (30) gets transformed into
a quarter circle. Namely, we obtain a prefactor of $-i\pi /2$ rather than $%
-i\pi $. Thus

\begin{equation}
ImS_{out}=\pi \omega r_{0}.  \label{31n}
\end{equation}

Similarly, we can obtain the imaginary part of action for the ingoing
particles as

\begin{equation}
ImS_{in}=-\pi \omega r_{0},  \label{32n}
\end{equation}

\bigskip so that from Eq. (15) we have

\begin{equation}
ImS=2\pi \omega r_{0}.  \label{33n}
\end{equation}

This result is in agreement with Eq. (23), and it leads to the conventional
Hawking temperature (8). In short, the failure of the PWTM in revealing
non-thermal radiation is proven also in the ICs.

\section{Conclusion}

In this article, it has been shown that the original PWTM method can not
convert isothermal HR to a non-thermal radiation. In particular, we have
used the LDBH, which radiates isothermally. In order to use the PWTM, the\
PGCs and ICs, which are two well-behaved coordinate systems, have been
chosen. In both coordinate systems, it has been straightforwardly shown
that, in spite of the energy conservation being taken into account, the pure
thermal character of the HR does not modify. Namely, the original PWTM does not resolve
the information loss paradox in the LDBH spacetime. Hence, it is our belief that seeking an alternative model to the PWTM, beside the work of \cite{Sakalli11}, which can produce the non-thermal radiation from the LDBH will be useful in the information theory. This is going to be our next problem in the near future. 

\section{Acknowledgement}
We thank the anonymous referee and editor for their valuable and constructive suggestions.


\begin{thebibliography}{99}
\bibitem{Hawking} S.W. Hawking, Commun. Math. Phys. \textbf{43}, 199 (1975); 
\textbf{46}, 206 (1976), Erratum.

\bibitem{ILP} V. Balasubramanian and B. Czech, Class. Quantum Grav. \textbf{%
28}, 163001 (2011).

\bibitem{PW} M.K. Parikh and F. Wilczek, Phys. Rev. Lett. \textbf{85}, 5042
(2000).

\bibitem{WKB} E. Merzbacher, \textit{Quantum Mechanics}, (3rd ed., John
Wiley \& Sons, New York, 1998).

\bibitem{Parikh} M. K. Parikh, Int. J. Mod. Phys. D \textbf{13}, 2351 (2004).

\bibitem{Vanzo} L. Vanzo, G. Acquaviva and R. Di Criscienzo, Class. Quantum
Grav. \textbf{28}, 183001 (2011).

\bibitem{Park} M. Liu, L. Liu, J. Zhang, J. Lu and J. Lu, Gen. Relativ.
Grav. \textbf{44}, 3139 (2012).

\bibitem{QQJ} Q.Q. Jiang and S. Q. Wu, Phys. Lett. B \textbf{635}, 151
(2006).

\bibitem{Sakalli1}  H. Pasaoglu, I. Sakalli, Int.J.Theor.Phys. \textbf{48}, 3517 (2009) .



\bibitem{Clement} G. Cl\'{e}ment, D. Gal'tsov and C. Leygnac, Phys. Rev. D 
\textbf{67}, 024012 (2003).

\bibitem{Fabris} G. Cl\'{e}ment, J.C. Fabris and G.T. Marques, Phys. Lett. B 
\textbf{651}, 54 (2007).

\bibitem{Marques} J.C. Fabris and G.T. Marques, Eur. Phys. J. C \textbf{72},
2214 (2012).

\bibitem{MSH} S. Mazharimousavi, I. Sakalli and M. Halilsoy, Phys. Lett. B 
\textbf{672}, 177 (2009).
\bibitem{Sakalli11} I. Sakalli, M. Halilsoy and H. Pasaoglu, Int. Jour.
Theor. Phys. \textbf{50}, 3212 (2011).


\bibitem{Sakalli2} I. Sakalli, M. Halilsoy and H. Pasaoglu, Astrophys. Space
Sci. \textbf{340}, 155 (2012).

\bibitem{ClemLey} G. Cl\'{e}ment, C. Leygnac, Phys. Rev. D \textbf{70},
084018 (2004).

\bibitem{BrownYork} J.D. Brown and J.W. York, Phys. Rev. D \textbf{47}, 1407
(1993).

\bibitem{Wald} R.M. Wald, \textit{General Relativity} (The University of
Chicago Press, Chicago and London, 1984).

\bibitem{Painleve} P. Painlev\'{e}, C. R. Acad. Sci. (Paris) \textbf{173},
677 (1921).

\bibitem{Gullstrand} A. Gullstrand, Arkiv. Mat. Astron. Fys. \textbf{16 (8)}%
, 1 (1922).

\bibitem{Hartle} J.B. Hartle and S.W. Hawking, Phys. Rev. D \textbf{13},
2188 (1976).

\bibitem{Wang} F.J. Wang, Y.X. Gui and C.R. Ma, Phys. Lett. B 650, \textbf{%
317} (2007).

\bibitem{Mathews} J.H. Mathews and R.W. Howell, \textit{Complex Analysis for
Mathematics and Engineering}, (6th ed., Jones and Bartlett Publishers,
London, 2012).

\bibitem{Landau} L.D. Landau and E.M. Lifshitz, T\textit{he Classical Theory
of Field}, (Pergamon, London, 1975).

\bibitem{SM} I. Sakalli and S.F. Mirekhtiary, Jour. Exp. Theor. Phys. 
\textbf{117}, 656 (2013).

\bibitem{Akhmedov} E.T. Akhmedov, V. Akhmedova, D. Singleton, Phys. Lett. B
642 (2006).

\bibitem{Chatterjee} B. Chatterjee and P. Mitra, Gen. Relativ. Grav. \textbf{%
44}, 2365 (2012).
\end{thebibliography}
\end{document}